\documentclass[%
 reprint,
 superscriptaddress,
 amsmath,amssymb,
 aps,
prmater,
floatfix
]{revtex4-2}

\usepackage{graphicx}
\usepackage{dcolumn}
\usepackage{bm}
\usepackage{upgreek}
\usepackage{comment}

\begin{document}

\AtEndEnvironment{thebibliography}{
\bibitem{data}
       C.~Brennan-Rich, S.~M.~Collins, S.~Micklethwaite, Z.~Aslam, T.~Almeida, S.~McVitie, R.~M.~Drummond-Brydson, and C.~H.~Marrows, Research Data Leeds, https://doi.org/TBC, 2025
}

\title{Structural and magnetic properties of co-sputtered epitaxial Fe-Sn kagome thin films}

\author{Callum~Brennan-Rich}
\affiliation{School of Physics and Astronomy, University of Leeds, Leeds LS2 9JT, United Kingdom}
 
\author{Sean~M.~Collins}%
\affiliation{School of Chemical and Process Engineering, University of Leeds, Leeds LS2 9JT, United Kingdom}
\affiliation{School of Chemistry, University of Leeds, Leeds LS2 9JT, United Kingdom}
\affiliation{Department of Materials, Royal School of Mines, Imperial College London, London SW7 2AZ, United Kingdom}

\author{Stuart~Micklethwaite}%
\affiliation{School of Chemical and Process Engineering, University of Leeds, Leeds LS2 9JT, United Kingdom}

\author{Zabeada~Aslam}%
\affiliation{School of Chemical and Process Engineering, University of Leeds, Leeds LS2 9JT, United Kingdom}

\author{Trevor~Almeida}%
\affiliation{School of Physics and Astronomy, University of Glasgow, Glasgow, G12 8QQ, United Kingdom}

\author{Stephen~McVitie}%
\affiliation{School of Physics and Astronomy, University of Glasgow, Glasgow, G12 8QQ, United Kingdom}

\author{Rik~M.~Drummond-Brydson}%
\affiliation{School of Chemical and Process Engineering, University of Leeds, Leeds LS2 9JT, United Kingdom}

\author{Christopher~H.~Marrows}
\email{c.h.marrows@leeds.ac.uk}
\affiliation{School of Physics and Astronomy, University of Leeds, Leeds LS2 9JT, United Kingdom}

\date{\today}

\begin{abstract}

In recent years the intermetallic alloys of Fe and Sn have gained significant interest due to a rich variety of magnetic properties present in these materials. The crystal Fe$_3$Sn$_2$ is a frustrated ferromagnet, while the crystallographically similar FeSn, which differs only by the stacking sequence of its Fe-containining kagome and stanene layers, is an antiferromagnet. Thin-film growth techniques such as magnetron sputtering allow for these different stoichiometric compositions to be grown through adjustments of the rate of deposition of the individual Fe and Sn sources, while all other conditions remain constant. Here, we report the production of high quality epitaxial thin films of Fe$_3$Sn$_2$ and FeSn on sapphire with a Pt seed layer, as well as a mixed phase containing intergrowths of both crystals, all of which we have characterized using both X-ray and four-dimensional scanning transmission electron microscopy (4D-STEM) methods. The resulting crystallographic phase content is compared to the results of magnetization measurements, with correspondence between the predicted ferromagnetic phase content and the resulting magnetization. Further magnetic properties of these films can then also be compared, leading to the discovery of a unique behavior in the temperature dependent coercivity within highly mixed phase alloys, a feature that is absent in either pure Fe$_3$Sn$_2$ and FeSn.

\end{abstract}

\maketitle


\section{\label{sec:Introduction}Introduction}

The five intermetallic alloys of Fe and Sn (FeSn$_2$, FeSn, Fe$_3$Sn$_2$, Fe$_5$Sn$_3$, and Fe$_3$Sn) have been the object of study for nearly a century \cite{Ehret}. Measurements have revealed a rich variety of magnetic properties and distinct differences between these alloys \cite{Trumpy}. In particular, recent work has focused on the two alloys FeSn \cite{Inoue, Sales, Li_FeSn, Satake} and Fe$_3$Sn$_2$ \cite{Zhang, Cheng, Kida, Wang_AHE} for their novel magnetic and electronic properties that arise from the kagome structure formed by the Fe sites in the Fe-containing atomic planes. These properties include the observation of room-temperature skyrmions stabilized by kagome-induced frustration in Fe$_3$Sn$_2$ \cite{Ren,Hou_stripe} that can have their chirality switched by a current pulse \cite{Hou}. The same kagome structure gives rise to exotic features in the electronic structure such as Dirac fermions and flat bands \cite{Ye,Kang}, which leads to non-Fermi liquid behavior in Fe$_3$Sn$_2$. The interlinking of magnetism and electronic structure in Fe$_3$Sn$_2$ comes about through the presence of Weyl nodes that can be moved in momentum space by changing the magnetization direction \cite{Yao18,Ren22}.

\begin{figure}[t]
\includegraphics[width=7.5cm]{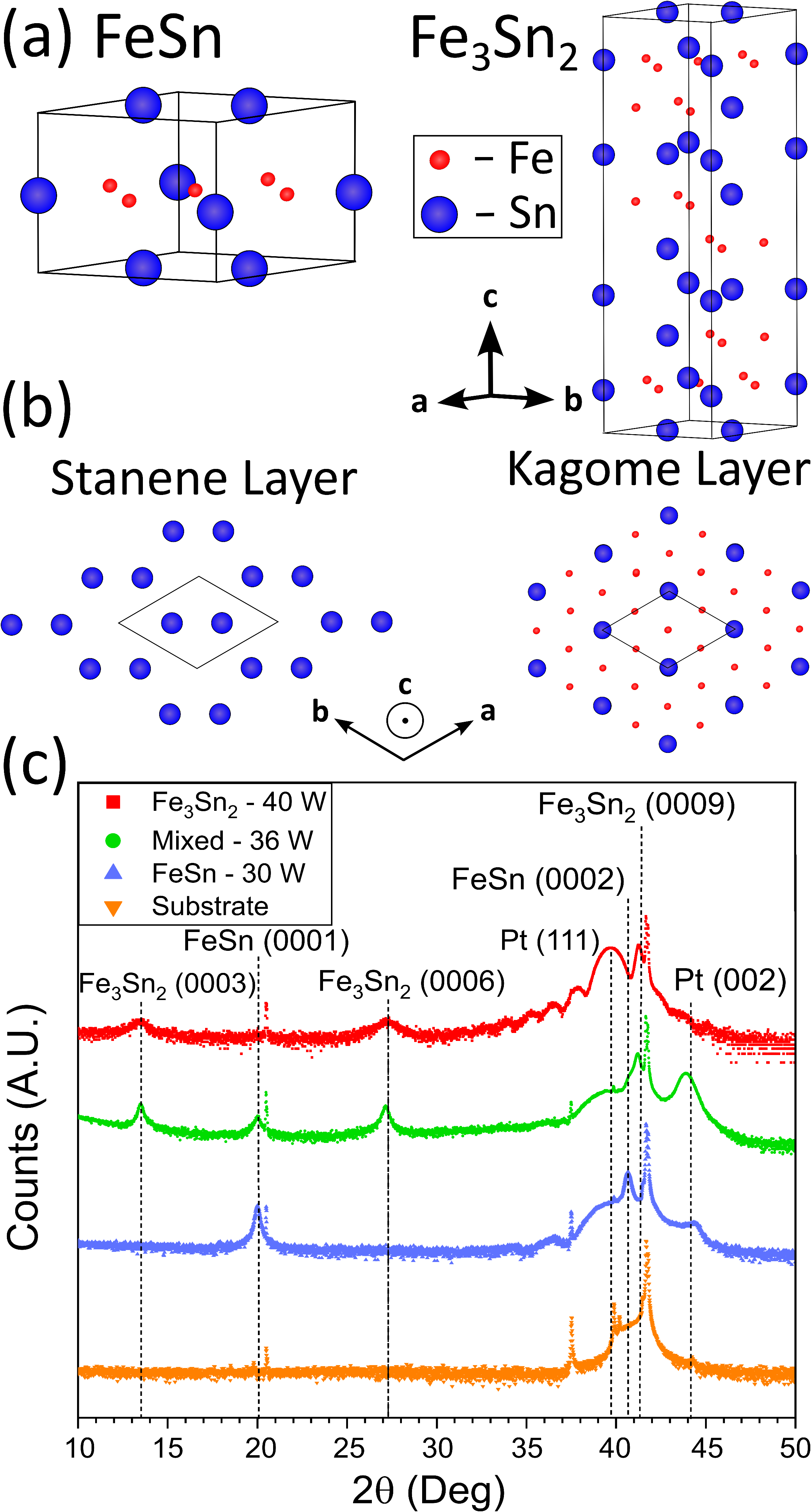}
\caption{\label{fig:Fig.1} Crystal structures of FeSn and Fe$_3$Sn$_2$. (a) Unit cells of the antiferromagnet FeSn and the frustrated ferromagnet Fe$_3$Sn$_2$ crystal. (b) Schematic of the Sn$_2$ stanene layer and the Fe$_3$Sn kagome layer. (c) 2$\theta-\omega$ XRD pattern of three co-sputtered thin films and one clean sapphire substrate. The sapphire (0006) peak seen at 41.68$^{\circ}$ on all patterns, as well as system peaks from the XRD equipment at 20.8$^{\circ}$ and 37.7$^{\circ}$. Films were grown with different growth power of the Fe gun, 40~W, 36~W, and 30~W. Dotted lines indicate the calculated angles for the labeled reflections simulated using the VESTA software \cite{VESTA} from the unit cells provided by Giefers and Nicol \cite{Giefers}.}
\end{figure}

These two intermetallic compounds have very closely related crystal structures, formed of two layers. The first is a pure Sn layer arranged in a honeycomb of hexagons (when isolated, this individual layer is sometimes called a ``stanene layer'' as an analogy to graphene \cite{Saxena}). The second layer has the chemical composition Fe$_3$Sn comprising Sn atoms at the corners of each cell of the repeating rhombus and an Fe atom in between each pair of Sn atoms (rhombus edge centers and rhombus center) \cite{Ghimire}. This results in the Fe atoms forming a kagome structure, consisting of regular triangles separated by hexagons as shown in Fig.~\ref{fig:Fig.1}(b). FeSn alternates between these two layers \cite{Nial} and is a hexagonal crystal with the $P6/mmm$ space group and lattice constants $a = b = 5.298$~\AA\ and $c = 4.448$~\AA\ \cite{Giefers}. Since the Sn atoms have no net magnetic moment, the magnetic properties of the material are completely determined by the spins in the Fe$_3$Sn and their interactions. For FeSn all moments within a given Fe$_3$Sn plane couple ferromagnetically, but moments between planes couple antiferromagnetically \cite{Hartmann}. Consequently, FeSn is an A-type antiferromagnet, and it exhibits a N\'eel temperature of 370~K \cite{Ericsson}.

Fe$_3$Sn$_2$, rather than just having alternating stanene and Fe$_3$Sn layers, has a more complicated layered structure with each stanene layer separated by two layers of Fe$_3$Sn. In turn, between the adjacent Fe$_3$Sn layers, each successive  Fe$_3$Sn layer is displaced by the stacking vector $\begin{pmatrix} 1/3&1/3&h \end{pmatrix}^T$ (in fractional lattice coordinates \cite{Roques}, where $h$ is the separation between layers in those coordinates). This lateral offset in the basal plane reduces the symmetry and means that Fe$_3$Sn$_2$ is described by a rhombohedral lattice (trigonal crystal system) rather than a hexagonal lattice (hexagonal crystal system). As such, Fe$_3$Sn$_2$ crystallizes in the space group $R\bar{3}m$ with lattice constants $a = b = 5.340$~\AA\ and $c = 19.797$~\AA\ \cite{Giefers}. 

Due to the bilayer of Fe$_3$Sn within its structure, Fe$_3$Sn$_2$ is a ferromagnet and exhibits a Curie temperature of about 657~K \cite{LeCaer}. In bulk form, there is a temperature-dependent reorientation of these moments: At room temperature, the moments are oriented along the $c$-plane, but below 60~K the moments change  to be oriented almost entirely in the basal plane\cite{Malaman,Haggstrom}. We note that this behavior will be affected also by shape effects in a thin film. Through examining the irreducible representation for the $R\bar{3}m$ group, Fenner et al. showed that this reorientation produces a frustrated arrangement of the spins \cite{Fenner}, rather than a collinear arrangement as previously expected.

Fe$_3$Sn$_2$ has been suggested as a possible candidate to host non-chiral magnetic skyrmions for data storage \cite{Periero,Ren,Hou} while FeSn has shown magnon-orbital coupling with potential uses in quantum information devices \cite{Kang}. Thin films, as opposed to bulk samples, are usually the way that these novel properties can find practical applications in spintronics, magnetic sensing, and data storage \cite{Sakthinathan}, and so understanding the properties of the thin films of these alloys and how to prepare them is essential for their future development. Epilayers of FeSn have previously been grown by molecular beam epitaxy \cite{Inoue,Hong}. This method has also been used for Fe$_3$Sn$_2$ \cite{Cheng,Ren22}, as well as facing target sputtering \cite{Zhang} and co-sputtering \cite{Khadkha}.

Here we demonstrate that the production of high quality epitaxial thin films of Fe$_3$Sn$_2$ and FeSn are possible within the same growth system through the use of sputter deposition on sapphire substrates using a Pt seed layer. Tuning the growth of one phase or the other can be controlled through stoichiometric changes during deposition due to the crystal structures of the phases being so similar, differing only in stacking sequence. Their characterization by high resolution and four dimensional scanning transmission electron microscopy (HRSTEM and 4D-STEM) reveals a much richer picture of the phase content than X-ray diffraction alone, including the distribution of the phase structure within the films. The basic magnetic properties of predominant Fe$_3$Sn$_2$ and FeSn samples match expectations based on their phase composition deduced from electron microscopy, although a film revealed to contain a mixture of both phases demonstrates hitherto unreported behavior in its coercivity vs temperature measurements, highlighting the importance of being able to precisely tune the growth parameters to predominantly produce the phase that is desired. 

\section{\label{sec:Methods}Materials and Methods}

All films in this work were grown using sputter deposition using a pair of co-sputtering magnetron sources. The substrate of $c$-plane sapphire was chosen as it is well established as a substrate for the growth of epitaxial Pt \cite{Farrow}. The epitaxial Pt in turn lattice matches well with the a-b spacing of Fe$_3$Sn$_2$ and FeSn. Large 40~mm diameter substrate wafers were first scored by a wafer saw before being marked to indicate the $m$-plane direction and broken by hand into 8~mm squares. All substrates were sonicated in acetone for five minutes to dissolve organic species adhering to the surface. Next, the substrates were sonicated  in isopropanol for another five minutes to remove any acetone residue. After drying in air with a hand-squeeze air pump, the substrates were visually inspected before being considered clean enough to enter the deposition system.

After the pump-down process was completed to $\leq~10^{-6}$~mbar, the heater elements were activated with a ramp rate of 50~\textdegree C per hour and a target of 500~\textdegree C. The substrates were then left for over 12 hours to allow the substrate temperature to stabilize at 500~\textdegree C. Typical chamber pressures at this temperature were in high $10^{-7}$~mbar range, which reduced to low $10^{-7}$ or high $10^{-8}$~mbar when the Meissner trap was activated. The Meissner trap was left to run for an hour with internal gas partial pressures monitored by a mass spectrometer. Water pressures could typically be reduced from a 90\% contribution of total pressure to 30\% (approximately equal to the pressures of N$_2$, O$_2$ and CO$_2$ present). Each material target was pre-sputtered for fifteen minutes to remove any potential oxide that might have formed on its surface. Growth rates were determined from X-ray reflectivity thickness measurements of calibration films. 

The Pt seed layers were then grown with a power of 12~W which resulted in a rate of about 1.3~\AA/s to grow 5~nm Pt seed layers. The Ar pressure during growth was $5 \times 10^{-3}$~mbar. These parameters were selected based on previous evidence of high-quality epitaxial growth \cite{Rodgers}. 

With these seed layers deposited, the substrate temperature was then reduced to 450~\textdegree C over the next fifteen minutes. Once this temperature was reached, the system was allowed to sit for an extra hour to ensure stability before the main growths were started. The Fe and Sn magnetron guns were simultaneously turned on with the Sn gun having a power of 10~W and the Fe varying depending on stoichiometry desired but with a typical power of about 35~W. The Ar pressure during growth was $5 \times 10^{-3}$~mbar. After completing the growths, the system was set to cool to room temperature at 200~\textdegree C an hour. Finally a Pt capping layer for the film was grown to protect the films from oxidation before the chamber was vented and the samples removed. 

\begin{table*}[bt]
\caption{\label{tab:table_struct_params}Structural parameters of the three samples used in this study, as determined by 4D-STEM analysis. The interplanar spacing for the (0001) direction is given since the lattice parameter $c$ is ill-defined for the Mixed sample, which does not possess a single crystal lattice; it is $d = c/9$ for Fe$_3$Sn$_2$ and $d = c/2$ for FeSn.}
\begin{ruledtabular}
\begin{tabular}{ccccc}
 Sample & Lattice Parameter $a=b$ (\AA)  & Interplanar Spacing $d$ (\AA) & Fe$_3$Sn$_2$ Content (\%)
& FeSn Content (\%) \\ \hline
 Fe$_3$Sn$_2$& $5.54 \pm 0.02$ & $2.215 \pm 0.001$ & $97.1 \pm 0.5$ & $2.9 \pm 0.5$ \\
 Mixed & $5.54 \pm 0.02$ & $2.187 \pm 0.001$ & $68 \pm 2$ & $33 \pm 2$ \\
 FeSn & $5.50 \pm 0.02$ & $2.184 \pm 0.001$ & $10.3 \pm 0.6$ & $89.7 \pm 0.6$ \\
\end{tabular}
\end{ruledtabular}
\end{table*}

\begin{figure}[t]
\includegraphics[width=8cm]{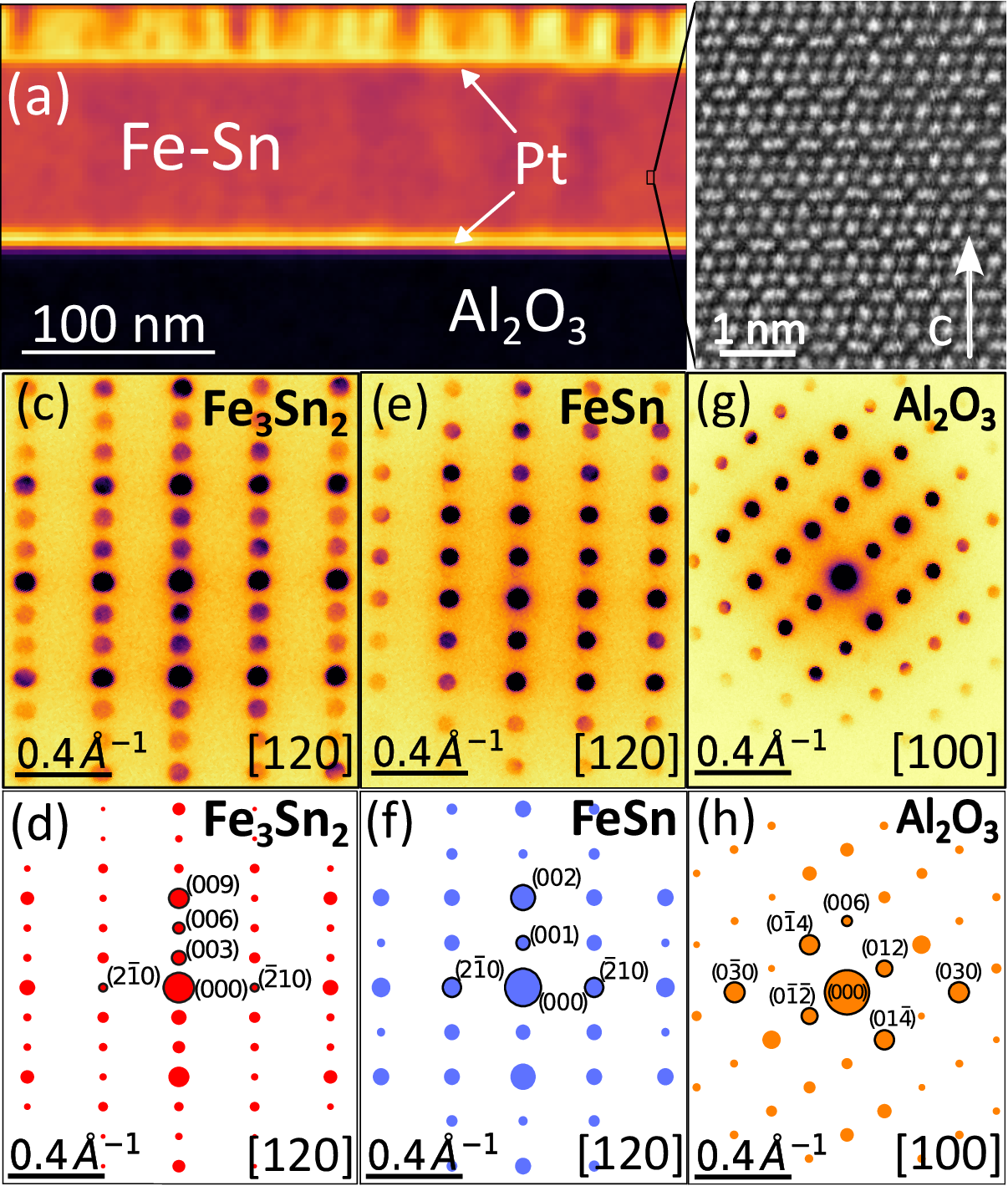}
\caption{\label{fig:Fig.2} STEM results: (a) HAADF image of cross section of the mixed film, with an expanded region explored with (b) high resolution aberration corrected dark field STEM revealing the individual atomic layers. The contrast between the low atomic number ($Z$) sapphire (Al$_2$O$_3$) substrate and high $Z$ Pt of the seed and capping layer are clear. Also shown (c-h) are the nanobeam electron diffraction (NBED) patterns as well as simulated NBED patterns used for orientation and indexing. These correspond to (c,d) Fe$_3$Sn$_2$, (e,f) FeSn, and (g,h) for the sapphire substrate.}
\end{figure}

4D-STEM experiments were carried out using the TESCAN Tensor system at the University of Leeds, a dedicated STEM equipped with a Schottky field emission gun electron source operated at 100 kV, precession optics, a Dectris Quadro counting-type electron detector, and detectors for X-ray energy dispersive spectroscopy (EDS). Additional high resolution STEM was carried out using an aberration-corrected JEOL ARM200CF microscope at the University of Glasgow. Cross section lamellae were milled using a focused ion beam (FIB) from all three films down to a thickness of between 60 and 70 nm. The 4D-STEM data were analyzed using an automated template matching algorithm. This assigned a phase, either Fe$_3$Sn$_2$ or FeSn, by comparing the observed NBED diffraction at each pixel, such as those shown in Fig.~\ref{fig:Fig.2}(d) and (e), to a library of simulated diffraction patterns from the different crystallographic phases on a pixel-by-pixel basis \cite{Rauch}. The pixel size was 3.3~nm~$\times$~3.3~nm. The simulated NBED patterns are produced for a range of different zone axis orientations and then cross correlated to the experimental diffraction patterns in order to assign a correlation coefficient which indicates which crystallographic phase is the most likely to be present at that point in the lamella, so generating a phase map \cite{Cautaerts}. All the following process was performed using the Pyxem extension of the Python Hyperspy package \cite{HyperSpy}.

Magnetic measurements were performed on a Quantum Design MPMS 3 superconducting quantum interference device vibrating sample magnetometer (SQUID-VSM). Due to limitations of this instrument, measurements from 3~K to 400~K (`regular mode') and measurements between 300~K and 600~K (`oven mode') had to be performed using different sample holders. This meant that the samples had to be removed, reloaded, and recalibrated so that continuous measurement from 3~K to 600~K was not possible. However, the crossover region between 300~K and 400~K allowed for confirmation that the reloading had not had a significant effect on the measured magnetic values. 

\section{\label{sec:Results}Results and Discussion}

\subsection{Film crystallography}

Three 80~nm thick films were grown with this methodology, one with an Fe magnetron power of 40~W, one with an Fe magnetron power of 36~W, and one with an Fe magnetron power of 30~W. (Sn sputter power was fixed at 10~W.) Film thickness was confirmed through the use of X-ray reflectometry (XRR). These films will be labeled as Fe$_3$Sn$_2$, Mixed, and FeSn, respectively. Their measured structural properties are summarized in Table~\ref{tab:table_struct_params}; we shall explain in detail below how these were determined. Fig.~\ref{fig:Fig.1}(c) shows the X-ray diffraction (XRD) patterns of the three thin film samples. Both Fe$_3$Sn$_2$ and FeSn samples show reflections at the expected angles based on VESTA simulations from the crystallographic data from Giefers and Nicol \cite{Giefers}. The 30~W and 40~W samples show characteristic FeSn and Fe$_3$Sn$_2$ diffraction peaks, respectively. Meanwhile, the 36~W sample exhibits both sets of diffraction peaks. The expected growth orientation for the Fe-Sn with $c$-axis directed out-of-plane is observed with only $(000l)$ peaks present. The fact that the Mixed film shows both Fe$_3$Sn$_2$ and FeSn diffraction peaks indicates the presence of ordered growth of both crystallographic structures. The $c$-plane lattice parameter corresponds well with those previously reported \cite{Giefers}, with strains of 0.7\% for the Fe$_3$Sn$_2$ sample and 2\% for FeSn sample along the c-axis. It is worth noting that the Pt peak for the 40~W sample is especially well formed, good enough to exhibit Pendell\"{o}sung fringes away from the central diffraction peak at 40$^{\circ}$.

Fig.~\ref{fig:Fig.2}(a) shows a cross section taken from the Mixed film. The high-angle annular dark field (HAADF) image reveals the well formed interfaces between substrate, seed, and film layers. Direct measurement of these layers confirms the expected thicknesses from the calibrated sample growth rates and sample thicknesses measured through X-ray reflectivity. High resolution STEM, Fig.~\ref{fig:Fig.2}(b), confirms that the regularity of the atomic layers within these films. In this particular image, a region with the Fe$_3$Sn$_2$ stacking sequence is shown, with growth along the [0001] direction, which is in turn parallel to the [0001] direction of the sapphire substrate. Regions corresponding to Fe$_3$Sn$_2$ and FeSn can be more readily distinguished by using the nanobeam electron diffraction (NBED) patterns, shown in Fig.~\ref{fig:Fig.2}(c) and (e), respectively. NBED patterns are also shown for the sapphire substrate in Fig.~\ref{fig:Fig.2}(g). Simulated NBED patterns generated using Pyxem for Fe$_3$Sn$_2$, FeSn and sapphire are shown in Fig.~\ref{fig:Fig.2}(d), (f) and (h) to confirm the expected zone axis and indices for the NBED patterns.

\subsection{4D-STEM phase identification}

Crystallographic data for the Fe-Sn phases were initially taken from the work of Giefers and Nicol \cite{Giefers}. In order to optimize the template matching, adjustments to the simulated $a=b$ lattice constant were required with  an adjustment from $a = b = 5.344$~\AA\ in Giefers \cite{Giefers} to a lattice constant that matches with (111) Pt, $a = b = 5.54$~\AA\ for the Mixed and Fe$_3$Sn$_2$ sample, and 5.50~\AA\ for the FeSn sample. This would correspond to an in-plane tensile strain of approximately 3\% from the lattice constants reported in the literature. These $a=b$ lattice parameters are presented in the first results column of Table \ref{tab:table_struct_params}. The average interplanar spacing $d$ along the $c$-axis growth direction is also shown in the table, we present this quantity for easier comparison between the three films. It shows a slight contraction related to the epitaxial strain.

\begin{figure*}[t]
\includegraphics[width=16cm]{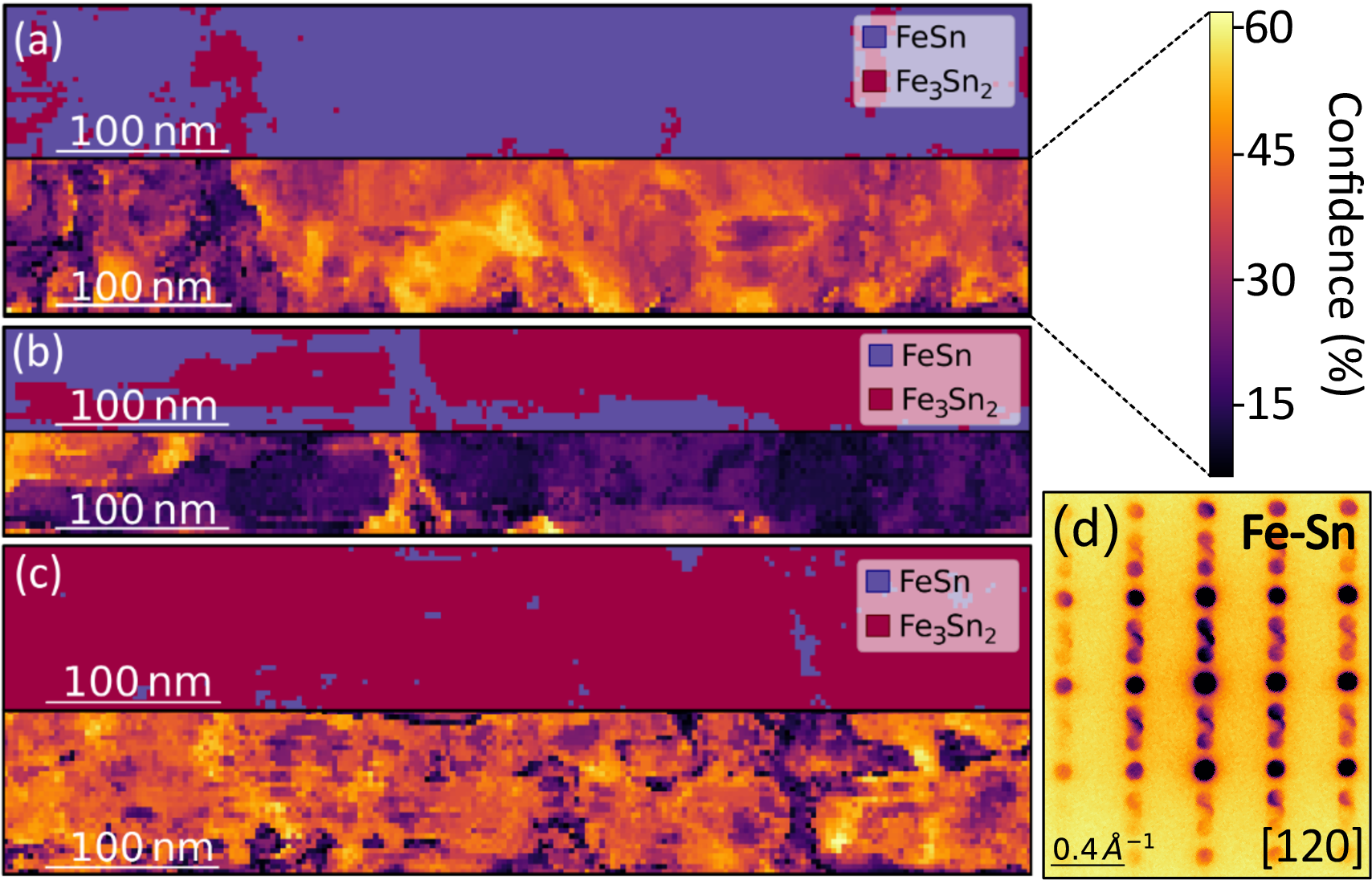}
\caption{\label{fig:Fig.3} Phase maps derived from 4D-STEM data for cross sections of: (a) the FeSn film, (b) the Mixed phase film, and (c) the Fe$_3$Sn$_2$ film.  Also shown below in each phase map are the associated confidence maps for each phase assignment with bright and dark representing high and low confidence respectively \cite{Rauch}. (d) shows the NBED pattern from a low confidence region of film, showing the blurred patterns for Fe$_3$Sn$_2$ and FeSn superimposed.}
\end{figure*}

The results of this 4D-STEM template matching process are shown in Fig.~\ref{fig:Fig.3}. Only the Fe-Sn film itself was analyzed in each case, and so only the regions between the Pt overcoat as well as the Pt seed and substrate are shown in the images. All three films are depicted as a binary colour map where each pixel has either been assigned red for Fe$_3$Sn$_2$ or blue for FeSn. The results quantify our qualitative expectations based on the X-ray diffraction results, with the film suspected of being purely FeSn sample being assigned $89.7 \pm 0.6$\% of its pixels as FeSn; the Mixed sample has $68 \pm 2$\% content of Fe$_3$Sn$_2$ and $33 \pm 2$\% FeSn; and the Fe$_3$Sn$_2$ sample as $97.1 \pm 0.5$\% Fe$_3$Sn$_2$. The full phase content results are given in Table~\ref{tab:table_struct_params}. 

Also shown in Fig.~\ref{fig:Fig.3} are the respective maps for the confidence coefficient for the assignments with bright indicating a relatively high confidence in the phase assignment and dark being a relatively low confidence, as indicated on the attached scalebar. It is generally true that the areas where the undesired phase exist within the two nominally phase pure films are associated with drops in confidence of assignment. Due to the thickness of the lamella and the 3.3 nm step size of the scanned electron probe, each pixel corresponds to the diffraction produced by a total interacting volume of about 800~nm$^3$. When an intergrowth of another phase has occurred, this volume of the FIB lamella will contain unit cells of both Fe$_3$Sn$_2$ and FeSn and so the resulting NBED pattern for that pixel will blur between the ideal Fe$_3$Sn$_2$ and FeSn patterns as shown in the NBED pattern presented in Fig.~\ref{fig:Fig.3}(d). For patterns of this type, the matching algorithm has to take a low confidence guess as to the phase assignment it selects. The generally lower confidence of the phase assignment for the Mixed film is the reason for the larger uncertainty in the phase percentages for this film. 

\subsection{Magnetic characterization}

\begin{figure}[t]
\includegraphics[width=6.5cm]{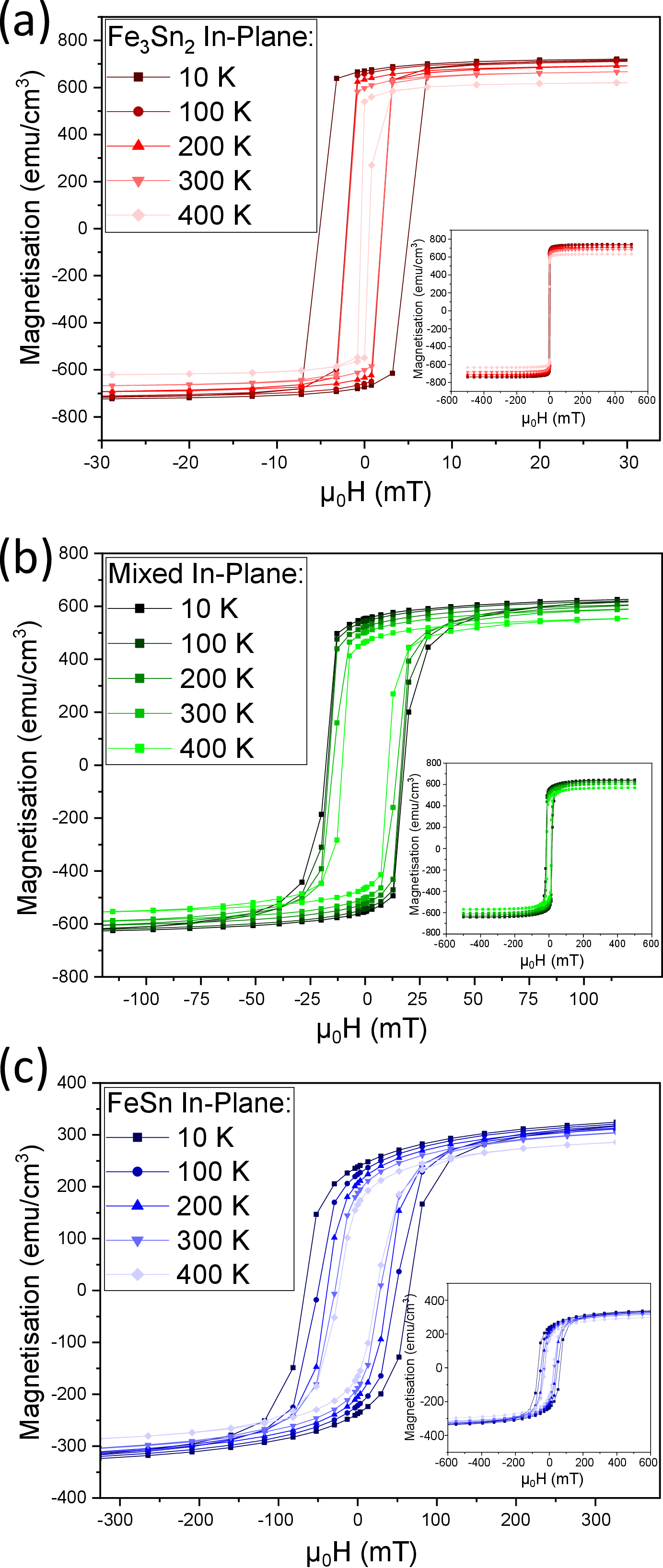}
\caption{\label{fig:Fig.4} In-plane hysteresis loops for the three films from 10~K to 400~K: (a) the Fe$_3$Sn$_2$ film; (b) the Mixed film; and (c) the FeSn film. Note the different field axis scales for the different panels. The inset in all cases shows the loops to full saturation at 600~mT.}
\end{figure}

Fig.~\ref{fig:Fig.4} shows the in-plane hysteresis loops, measured by SQUID-VSM, for the three samples over a range of temperatures. All three films were found to have an easy plane, with a hard axis out-of-plane, owing to shape anisotropy. The saturation field, $\mu_0 H_\mathrm{s}$, required to saturate the magnetization out-of-plane (loops not shown) was found to be $1010 \pm 30$~mT for the Fe$_3$Sn$_2$ film; $780 \pm 40$~mT for the Mixed film and $200 \pm 60$~mT for the FeSn film. 

In each graph shown in Fig.~\ref{fig:Fig.4}, the scale for the magnetic field axis has been chosen so that that the switching of the magnetization is visible. The inset in each of these graphs shows the full loops taken out to saturation. The saturation magnetization at 300~K is found to be $255 \pm \ 5$~emu/cm$^3$ for the FeSn sample, which is significantly smaller than the Mixed sample with a value of $616 \pm 8$~emu/cm$^3$. This value for the Mixed sample is in turn smaller than the Fe$_3$Sn$_2$ sample with $835 \pm 9$~emu/cm$^3$. This measured magnetisation of Fe$_3$Sn$_2$ is larger than those previously reported, with typical saturation magnetisation in the literature reported to be in the region of 600 to 700 emu/cm$^{3}$ \cite{Khan,Khadkha}, perhaps indicating the presence of Fe impurities raising the overall magnetisation. 

All three samples were found to exhibit some degree of ferromagnetism, which is expected due to the 4DSTEM analysis revealing the presence of some Fe$_3$Sn$_2$, even in the nominally FeSn sample. Nevertheless, since FeSn is expected to be antiferromagnetic with zero magnetization, a natural comparison is the percentage content predicted from the phase mapping in Fig.~\ref{fig:Fig.3} and the resulting ratios of magnetization in Fig.~\ref{fig:Fig.4}. The ratio of Fe$_3$Sn$_2$ content between the Mixed sample and the Fe$_3$Sn$_2$ sample extracted from the phase map is $0.70 \pm 0.02$, while the ratio of their saturation magnetizations is in good agreement at $0.73 \pm 0.02$. However, the ratio of Fe$_3$Sn$_2$ phase content between the FeSn and Fe$_3$Sn$_2$ samples is $0.106 \pm 0.008$ while the ratio of the magnetization is significantly higher with the FeSn film having $0.30 \pm 0.03$ of the magnetization of the Fe$_3$Sn$_2$ sample. A reasonable explanation for this comes from the observation of superimposed NBED patterns. Through multiple iterations of coding the template matching algorithm was programmed to be stricter on what was considered a ``good'' match to Fe$_3$Sn$_2$, until only patterns that were clearly Fe$_3$Sn$_2$ were being assigned as such. In a binary assignment, this inevitably means that blurred NBED patterns will get assigned more often to FeSn, though these areas do not truly represent fully compensated AFM regions, and so will still provide some net magnetization to the sample.

\begin{figure}[t]
\includegraphics[width=6.5cm]{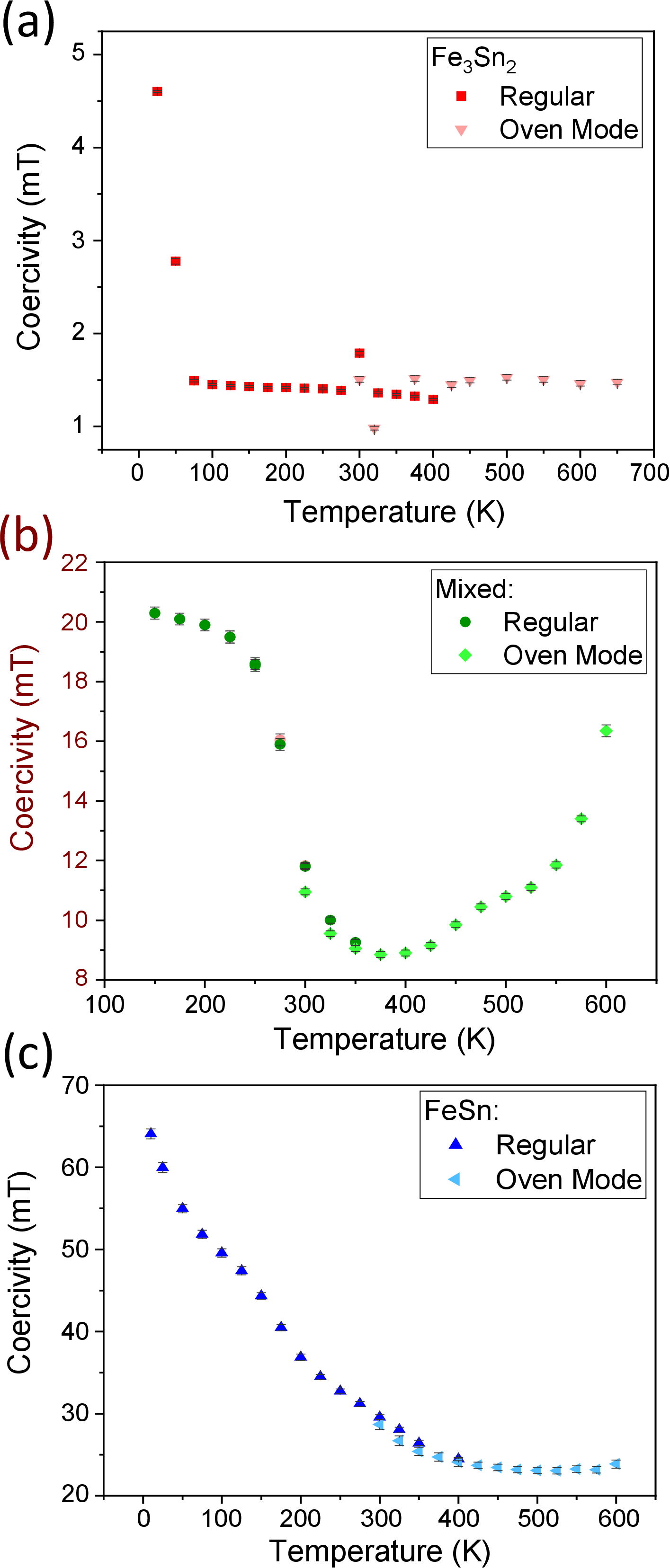}
\caption{\label{fig:Fig.5} Coercivity vs. temperature curves between 5 and 600 K for: (a) the Fe$_3$Sn$_2$ film; (b) the Mixed film; and (c) the FeSn film. The measurements taken with the SQUID-VSM in regular and oven mode are indicated. Error bars indicate individual uncertainties associated with measuring a particular coercive field from the hysteresis loop. The Fe$_3$Sn$_2$ film and FeSn film show an expected decrease in coercivity with increasing temperature, but the mixed film demonstrates a minimum in coercivity at 375~K.}
\end{figure}

The value of the coercivity also changes significantly between the samples. At 300~K the Fe$_3$Sn$_2$ sample has a coercivity, $\mu_0H_\text{c}$, of $1.4 \pm 0.2$~mT, the Mixed sample has a coercivity of $11.2 \pm 0.5$~mT; and the FeSn sample has a coercivity of $29.0 \pm 0.8$~mT. Fig.~\ref{fig:Fig.5} shows the change in coercivity as the the temperature is varied from 5~K to 600~K. The changeover between the low temperature `regular mode' and the high temperature `oven mode' of the magnetometer is indicated. Both Fe$_3$Sn$_2$ and FeSn sample show the expected reduction in coercivity at higher temperatures, where reversal is assisted by thermal activation. The drop in coercivity is remarkably steep at the lowest temperatures for the Fe$_3$Sn$_2$ film. 


However, the Mixed sample shows a completely different and very unusual behavior. There is an initial drop in coercivity occurring as temperature rises until a minimum in coercivity is reached at 375~K. Increasing temperature above that point sees the coercivity rise further with a slight inflection occurring at 475 K before a very rapid rise in coercivity up to 600 K which was the highest temperature measured. This behavior is highly repeatable and so was unlikely to be associated with any permanent crystallographic change occurring in the Mixed sample at elevated temperatures.

One possible explanation for this unusual behavior could be the interaction between regions of ferromagnetic Fe$_3$Sn$_2$ and antiferromagnetic FeSn, which gives rise to a set of phenomena referred to as exchange bias \cite{Nogues}, which can include changes to the coercivity as well as the more usually observed shift in switching along the field axis. The work of Zhang~\textit{et al}. \cite{ZhangCoercivity} has associated the ferromagnetic to antiferromagnetic boundary coupling with an increase in coercivity, which usually peaks around the temperature where the antiferromagnet becomes disordered. However, in the case of the Mixed sample here, it is a minimum in coercivity that corresponds closely to the N\'eel temperature of 370~K for FeSn \cite{Ericsson}.  

\section{\label{sec:Conclusions}Conclusions}

We have grown epitaxial Fe-Sn alloy films by co-sputtering onto sapphire substrates with a Pt seed layer. We have studied three films in detail: one that shows only XRD peaks for Fe$_3$Sn$_2$, one showing only FeSn peaks, and one exhibiting both sets of peaks (`Mixed'). X-ray crystallography, electron microscopy and magnetometry have been shown to work well in tandem to characterize these three films. XRD in particular is a reliable and fast technique to highlight which films are worth examining in closer detail with SQUID-VSM and STEM measurements. The phase content deduced through the 4D-STEM pattern matching process shows that the two nominally phase pure films contain at least $\sim 90$\% of the predominant phase.

The Fe$_3$Sn$_2$ has a magnetization measured by SQUID-VSM close to other literature values for thin films and bulk crystals. The magnetization of the Mixed sample is consistent with its Fe$_3$Sn$_2$ content as determined by 4DSTEM, with the FeSn sample being antiferromagnetic. The FeSn sample however has 30\% of the magnetisation of the Fe$_3$Sn$_2$ sample with only 11\% of the phase content, which is attributed to the simplicity of applying a binary phase assignment when each pixel may itself have some percentage contribution from both Fe$_3$Sn$_2$ and FeSn. A natural extension of this work would be to produce new template libraries that contain the blurred spot diffraction patterns and assign these as a third phase in the phase map. Another less sophisticated option would be to assign a mixed phase designation to any region of the film that has a confidence value in its Fe$_3$Sn$_2$ or FeSn assignment below some cut-off value.

The resulting magnetic properties of the film also seem to be highly dependent on the phase mixing, with the coercivity as a function of temperature showing an unexplained minimum. Potential origins and avenues of future study  are either an exchange bias interaction, or some result of unique domain wall motion that only occurs in the Mixed sample. Further investigation of the presence of skyrmions within these thin films, as well as the potential exotic transport properties produced by the kagome structure are also planned.

\section{\label{sec:Acknowledgments}Acknowledgments}

The authors would like to acknowledge and thank the discussions and technical support of A. Brown, G. Burnell, and M. Ali. This work was funded through a UK Engineering and Physical Sciences Research Council (EPSRC, 2597145) in the Bragg Centre for Materials Research at the University of Leeds. S.M.C. and R.M.D.-B. also acknowledge funding support from the EPSRC (EP/X040992/1).

\section*{\label{sec:DataAvailibility}Data Availability}

The data associated with this paper are openly available at the Research Data Leeds repository~\cite{data}.



\bibliography{FeSn}

\end{document}